%% file: main.tex
\documentclass[%
reprint,
superscriptaddress,
showkeys,
nobibnotes,
amsmath,amssymb,
aps,
prl,
]{revtex4-2}

\usepackage{graphicx}
\usepackage{dcolumn}
\usepackage{bm}
\usepackage{hyperref}
\usepackage{verbatim}
\usepackage[utf8]{inputenc}
\usepackage{layouts}
\usepackage{multirow}
\usepackage{lineno}
\usepackage[dvipsnames]{xcolor}
\usepackage{hhline} 
\usepackage[separate-uncertainty=true]{siunitx}
\usepackage[version=3]{mhchem} 
\usepackage{wrapfig}
\usepackage{braket}
\usepackage{float}
\usepackage{leftidx} 
\usepackage{amsmath,mathtools}
\usepackage{amssymb}
\usepackage{chngcntr}
\hypersetup{colorlinks=true,urlcolor=magenta,linkcolor=magenta,citecolor=cyan}

\newcommand{\beginsupplement}{%
  \clearpage
  \setcounter{section}{0}%
  \setcounter{subsection}{0}%
  \setcounter{figure}{0}%
  \setcounter{table}{0}%
  \setcounter{equation}{0}%
  \renewcommand{\thesection}{S\arabic{section}}%
  \renewcommand{\thesubsection}{S\arabic{section}.\arabic{subsection}}%
  \renewcommand{\thefigure}{S\arabic{figure}}%
  \renewcommand{\thetable}{S\arabic{table}}%
  \renewcommand{\theequation}{S\arabic{equation}}%
}

\newcommand{\maybeonecolumn}{%
  \ifdefined\onecolumngrid
    \onecolumngrid
  \else
    \ifdefined\onecolumn
      \onecolumn
    \fi
  \fi
}
\newcommand{\maybetwocolumn}{%
  \ifdefined\twocolumngrid
    \twocolumngrid
  \else
    \ifdefined\twocolumn
      \twocolumn
    \fi
  \fi
}

\begin{document}

\title{Single-Photon Detection in Few-Layer NbSe$_2$ Superconducting Nanowires}

\author{Lucio Zugliani}
\email{lucio.zugliani@tum.de}
\affiliation{%
    Walter Schottky Institute, TUM School of Computation, Information and Technology, and MCQST, Technical University of Munich, Munich, Germany
}%

\author{Alessandro Palermo}%
\affiliation{%
    Walter Schottky Institute, TUM School of Computation, Information and Technology, and MCQST, Technical University of Munich, Munich, Germany
}%

\author{Bianca Scaparra}%
\affiliation{%
    Walter Schottky Institute, TUM School of Computation, Information and Technology, and MCQST, Technical University of Munich, Munich, Germany
}%

\author{Aniket Patra}%
\affiliation{%
    Munich Quantum Instruments GmbH, Munich, Germany
}%

\author{Fabian Wietschorke}%
\affiliation{%
    Walter Schottky Institute, TUM School of Computation, Information and Technology, and MCQST, Technical University of Munich, Munich, Germany
}%

\author{Pietro Metuh}%
\affiliation{%
    Department of Electrical and Photonics Engineering, Technical University of Denmark, 2800 Kgs. Lyngby, Denmark
}%

\author{Athanasios Paralikis}%
\affiliation{%
 Department of Electrical and Photonics Engineering, Technical University of Denmark, 2800 Kgs. Lyngby, Denmark
}%


\author{Domenico De Fazio}%
\affiliation{%
    Department of Molecular Sciences and Nanosystems, Ca’ Foscari University of Venice, Venice, Italy
}%

\author{Christoph Kastl}%
\affiliation{%
    Walter Schottky Institute, TUM School of Natural Sciences, and MCQST, Technical University of Munich, Munich, Germany
}%

\author{Rasmus Flaschmann}%
\affiliation{%
    Walter Schottky Institute, TUM School of Computation, Information and Technology, and MCQST, Technical University of Munich, Munich, Germany
}%

\author{Battulga Munkhbat}%
\affiliation{%
    Department of Electrical and Photonics Engineering, Technical University of Denmark, 2800 Kgs. Lyngby, Denmark
}%

\author{Kai Müller}%
\affiliation{%
    Walter Schottky Institute, TUM School of Computation, Information and Technology, and MCQST, Technical University of Munich, Munich, Germany
}%

\author{Jonathan J. Finley}%
\email{jj.finley@tum.de}
\affiliation{%
    Walter Schottky Institute, TUM School of Natural Sciences, and MCQST, Technical University of Munich, Munich, Germany
}%

\author{Matteo Barbone}%
\email{matteo.barbone@wsi.tum.de}
\affiliation{%
    Walter Schottky Institute, TUM School of Computation, Information and Technology, and MCQST, Technical University of Munich, Munich, Germany
}%

\date{\today}

\begin{abstract}
Superconducting Nanowire Single-Photon Detectors (SNSPDs) are key building blocks for photonic quantum technologies due to their ability to detect single photons with ultra-high efficiency, low dark counts and fast temporal resolution. Superconducting materials exhibiting high uniformity, large absorption cross-section and atomic-scale thickness are desirable to extend single‑photon detection from the near‑infrared up to the terahertz regime, where existing material choices are especially constrained. Substrate independence would further open the way to integrate detectors onto functional materials and heterostructures, enhancing performance and enabling proximal read‑out of a wide range of individual excitations. Here, we top-down shape the prototypical two‑dimensional superconductor niobium diselenide (NbSe$_2$) into few-layer nanowires less than \SI{100}{\nano\metre} wide and demonstrate single‑photon detection at \SI{780}{\nano\metre} and \SI{1550}{\nano\metre}. At the same time, the dark-count rate remains below \SI{1}{\hertz} up to the switching current and we achieve a timing jitter below \SI{50}{\pico\second}. We use a diffusive hot‑spot model to estimate a theoretical cut‑off wavelength that surpasses the millimetre range. Our results open up routes toward quantum limited detectors integrated into quantum‑photonic circuits and quantum devices, with the potential for novel detection capabilities and unprecedented energy sensitivity.
\end{abstract}

\maketitle

\subsection*{Introduction}


SNSPDs combine unmatched high‑end performance across all key single‑photon detection indicators\cite{goltsman_2001, hadfield_2009}—from $>$\SI{99}{\percent} detector efficiency (DE)\cite{chang_2021}, to $<$\SI{5}{\pico\second} timing jitter\cite{korzh_2020}, $<$\SI{10}{\nano\second} recovery time\cite{cherednichenko_a2021}, and $<$\SI{1}{\hertz} dark‑count rates (DCR)\cite{taylor_2023}.
Due to such performance metrics, SNSPDs have recently found many near‑ideal applications in metrology, sensing and astronomy, and have cemented themselves as indispensable ally in optical quantum computing and quantum communications\cite{Natarajan_Tanner_Hadfield_2012, montblanch_2023}.

Development of material platforms has been at centre stage for quantum technologies\cite{acin_2018}, and SNSPDs are no exception. Niobium‑based superconductors (NbN, NbTiN) provide good timing resolution and fast recovery\cite{zichi_nbtin_2019, zugliani_engineering_2023}, while amorphous silicides (MoSi, WSi) offer enhanced sensitivity to longer wavelengths.\cite{kondo_wsi_1992, boswroth_mosi_2015, lita_mosi_2021, grotowski_mosi_2025}
Most recently, layered materials have emerged as a promising platform for quantum photonic technologies\cite{montblanch_2023}, with potential applications spanning from deterministic arrays of single-photon emitters\cite{palacios_2017}, to quantum light emitting diodes\cite{palacios_2016}, to novel spin-photon interfaces for qubits\cite{brotons_2019, stern_2024}, and analog quantum simulators\cite{tang_2020,kennes_2021}. 

Layered materials offer a unique combination of advantages for single-photon detection, such as atomic thickness and fast thermalization time\cite{Semenov_Engel_Hübers_Il’in_Siegel_2005} necessary for detecting low-energy excitations\cite{Semenov_Engel_Hübers_Il’in_Siegel_2005, Bulaevskii_Graf_Kogan_2012}; substrate independence, which allows for greater flexibility in device design, integration, and scope; as well as 
near-ideal uniformity due to sharp interfaces, chemically terminated surfaces, and multi-$\mu$m-scale single-crystal size, forecasting operational currents extremely close to the depairing currents as well as fast time resolution\cite{allmaras_intrinsic_2019}. Such advantages are especially appealing for high sensitivity, fast detection of long-wavelength photons up to the terahertz range\cite{vicarelli_micromechanical_2022,tredicucci_infrared_2024}, where available technologies remain limited\cite{taylor_2023}. Additionally, layered materials allow for the development of detectors integrated on-chip capable of sensing other degrees of freedom ranging from magnetic (spin) excitations to excitons, plasmons, phonons, and possibly even more exotic phases\cite{savary_quantum_2016}.

The technological barriers to using layered materials for single-photon detection\cite{orchin_2019} were only recently overcome in the high-temperature layered cuprate superconductor Bi$_2$Sr$_2$CaCu$_2$O$_{(8-\delta)}$\cite{charaev_2023} and in moiré superconductor magic-angle twisted bilayer graphene (MATBG)\cite{di_battista_2024}, showcasing the potential of layered materials for single-photon counting devices. However, no SNSPDs with conventional layered superconducting materials, which promises to offer a superior combination of near-atomic thickness, scalability, speed, and sensitivity to low energies, have been yet reported. 
Niobium diselenide (NbSe$_2$) is a van der Waals material with a temperature-dependent phase diagram including a normal metallic phase, hosting charge density waves\cite{xi_strongly_2015}. For temperatures below 10 K, it exhibits layer-dependent superconducting transition that persists down to the three-atom, $\sim$0.7 nm-thick monolayer limit\cite{cao_2015, xi_ising_2016}. 
NbSe$_2$ offers high absorption, $\sim$1 ps carrier thermalization\cite{wickramaratne_ising_2020} and can be processed using standard top-down nanofabrication techniques\cite{battulga_nbse2_fab_2023}. Moreover, it can be grown using scalable methods, such as vapor deposition at the wafer scale\cite{wang_high-quality_2017}, and benefits from integration with other van der Waals materials, such as hexagonal boron nitride (hBN). 
Few-layer NbSe$_2$ was used for multi-photon detection in the NIR\cite{orchin_2019} and terahertz\cite{shein_fundamental_2024}, and revealed hints of discrete-photon sensitivity when patterned into nanowires\cite{metuh2025}, but detecting single-photons remained elusive until now. 
Here, we report the  linear increase of the photon count rate under varying photon excitation power in a 7-layer-thick top-hBN encapsulated NbSe$_2$ superconducting nanowire. The count rate follows a Poisson distribution for single events under coherent (laser) excitation at \SI{780}{\nano\metre} and \SI{1550}{\nano\metre} with an incident power spanning multiple orders of magnitude, demonstrating single-photon detection\cite{goltsman_2001, semenov_statistic_2001}. 

\subsection*{Device fabrication and electrical behaviour}

\begin{figure*}[t]
\includegraphics[width=1\textwidth]{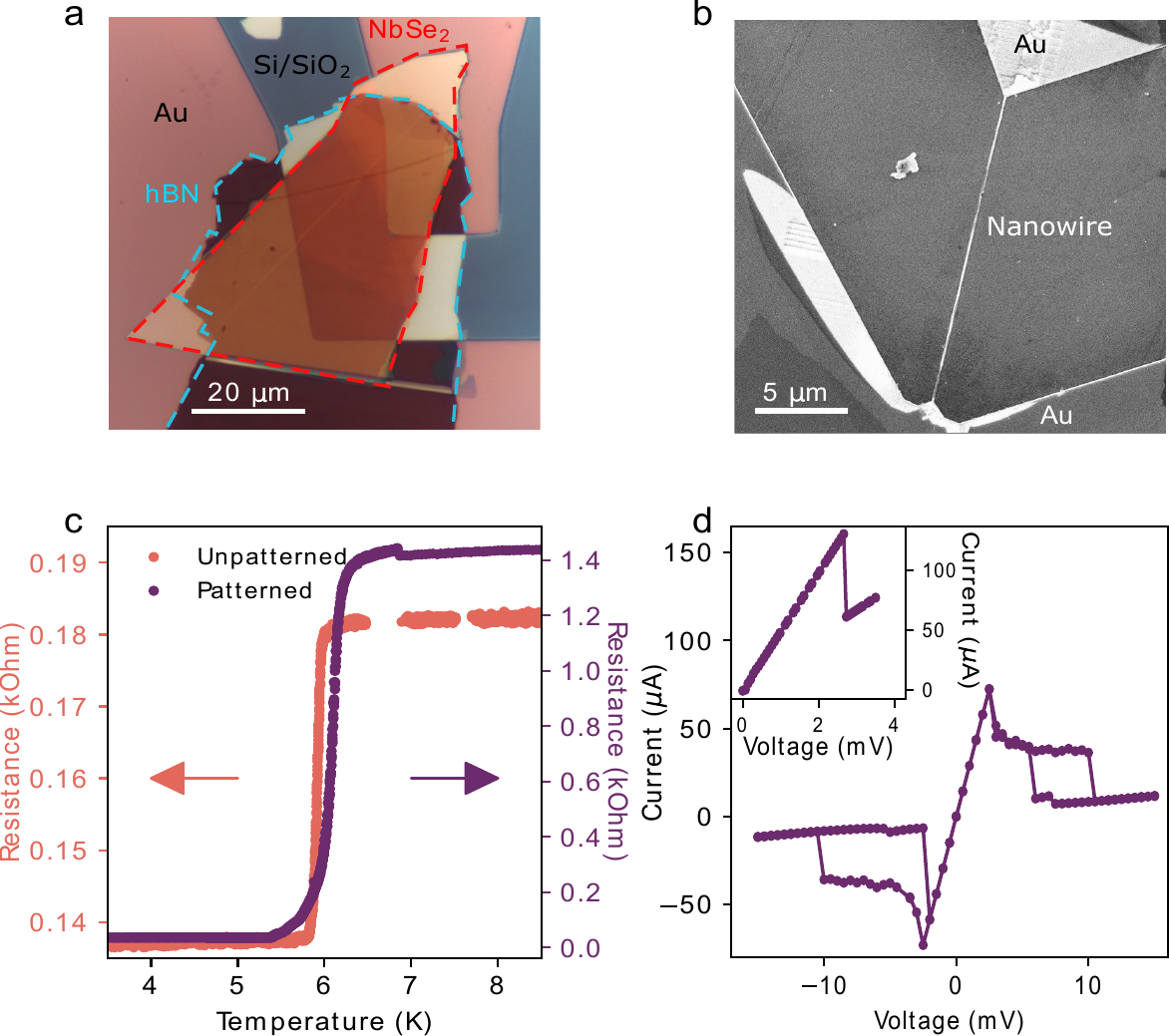}
\caption{\label{fig:device} $\textbf{Fabrication and cryogenic temperature electrical behaviour of NbSe$_2$-nanowire devices.}$
$\mathbf{a}$, Optical image of a hBN/NbSe$_2$-heterostructure transferred onto pre-patterned contacts (pink). $\mathbf{b}$, SEM image of the device after shaping the transport channel into a $\sim$\SI{100}{\nano\metre} nanowire. $\mathbf{c}$, Resistance as function of temperature shows superconducting transition of \SI{6}{\kelvin} for a seven-layer crystal before and after patterning. $\mathbf{d}$, I-V characteristics at \SI{1}{\kelvin} of the patterned device: in the positive regime, the switching current at \SI{2.4}{\milli\volt} is followed by an intermediate current state where superconductivity is broken and recovered up to \SI{10}{\milli\volt}, where the wire becomes fully resistive. The inset shows the I-V characteristics becoming a single, sharp transition from superconducting to resistive by adding a shunt resistor.}
\end{figure*}

Figure \ref{fig:device} summarizes the device fabrication flow. We assemble stacks of multilayer NbSe$_2$/hBN on oxidized silicon wafers according to conventional micromechanical cleavage, followed by dry-transfer processes\cite{castellanos_2014, purdie_2018}. The stacks are transferred on top of pre-patterned Ti/Au electrodes, with the NbSe$_2$ layer in contact with the metal (Fig. \ref{fig:device}a). Subsequently, we shape the heterostructures into \SI{60}-\SI{100}{\nano\metre} wide nanowires using standard nanofabrication techniques (Fig. \ref{fig:device}b). Further details on the fabrication process are included in the Methods Section. 

We then continue to study the two-point electrical transport behaviour of the devices at cryogenic temperatures (see Methods for details on the experimental setup). 
Figure \ref{fig:device}c shows typical superconducting transition temperature measurements of a selected device pre- and post-patterning of the nanowire (see Supplementary Information for transition temperatures of nanowires with different thicknesses). The critical temperature for both the un-patterned and patterned device is $\sim$$\SI{6.0}{\kelvin}$ and the superconducting transition occurs with a sharp transition $\sim$$\SI{0.5}{\kelvin}$ broad, which confirms the homogeneity of the flake and is in good agreement with previously reported values for both unpatterned \cite{cao_2015, xi_ising_2016} and patterned\cite{metuh2025} few-layer ($\leq10$) NbSe$_2$ flakes. 
After patterning the heterostructure to form a nanowire, we observe a change in the NbSe$_2$ film resistance given the narrower geometry of the patterned device, increasing to 1.40 k$\Omega$, translating to a resistivity $\rho\sim$6 $\mu\Omega\cdot$cm (sheet resistance $R_\square\sim$14 $\Omega / \square$). These values are more than an order of magnitude lower than conventional sputtered superconducting films\cite{korzh_2020} due to the homogeneity of the crystal. In contrast, the critical temperature remains minimally affected by top-down patterning. The residual resistance of $\sim$100 $\Omega$ below the superconducting transition is due to the contact resistance between NbSe$_2$ and the metal contacts as well as the wiring integrated in the setup. 


Figure \ref{fig:device}d shows the $I-V$ characteristics measured at $\sim$\SI{1}{\kelvin}, which allows to extract the switching current $I_{sw}$ corresponding to the superconducting-to-resistive transition. Initially, the nanowire is in a superconducting state. While increasing the voltage, we see an ohmic increase of the current governed by the contact and the wiring resistance of the setup. At about 2.4 mV the $I-V$ curve reaches a regime where the device is in a state of intermediate resistance. This is caused by the oscillation between superconducting and normal conducting state occurring because of the short recovery time of the device related to its low kinetic inductance\cite{kerman_latching}. Increasing the bias voltage further, we reach a second ohmic regime where the detector switches fully to the normal conducting state. This behaviour is also observed for reverse voltage. To remove the intermediate oscillatory behaviour and obtain a single clear phase transition, we connect a resistor in parallel to the device. The inset shows the $I-V$ characteristics recorded in such case. In this configuration, the current is redirected from the device to the resistor, decreasing Joule heating and therefore allowing a faster relaxation to the superconducting state. We note that in this case we observe a sharp transition between superconducting phase and normal conductivity, due to the decoupling of the electro-thermal components of the device.

\subsection*{Photoresponse characteristics}

\begin{figure*}
\includegraphics[width=1\textwidth]{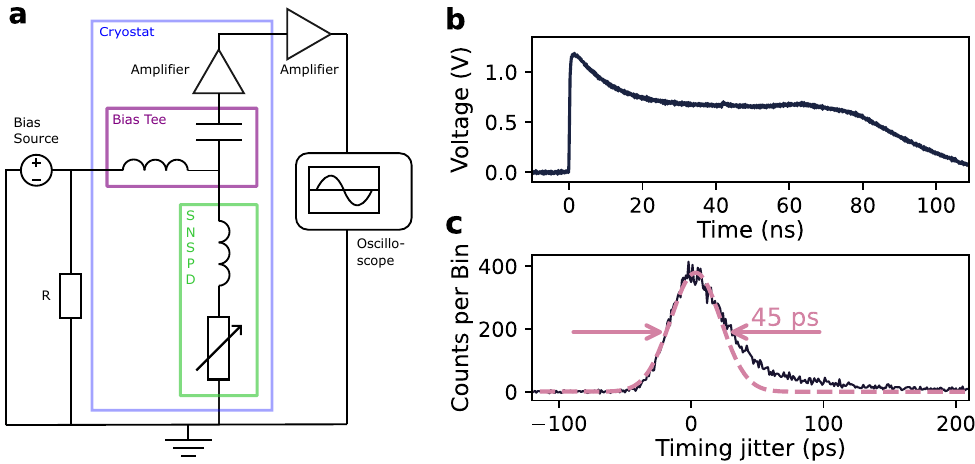}
\caption{\label{fig:pulse}$\textbf{Photoresponse characteristics of the detectors.}$ 
$\mathbf{a}$, Schematic of the electronic read-out circuit. A voltage source is used to supply a bias current to the detector. A parallel resistor $R$ is utilized to mitigate latching. The current is supplied to the detector via the DC arm of a bias-tee (purple) installed inside a cryostat working at \SI{4}{\kelvin} (blue). The detector (green) is represented by an inductance $L_{\mathrm{k}}$ in series with a variable resistor. When a photon is absorbed, the sudden increase in resistance generates a voltage pulse traveling through the AC arm of the cryogenic bias-tee (purple) and is consecutively amplified both at cryogenic temperature as well as at room temperature. The signal is then recorded via an oscilloscope. $\mathbf{b}$, Time dependent voltage pulse. The decay initially follows conventional single-exponential behaviour but then shows latching, where the superconducting state is not recovered until the current is redirected through the parallel resistor. 
$\mathbf{c}$, Timing jitter of the device with pulse shown in \textit{b}, where the gaussian fit returns a full-width-at-half-maximum value of \SI{45}{\pico\second}.}
\end{figure*}

We measure the photoresponse of the detectors in a cryogenic setup at $\sim$\SI{1}{\kelvin}, see Methods for details. 
Fig. \ref{fig:pulse}a shows a schematic of the read-out circuit. The source current flows through a bias-tee formed of a DC filter (capacitance) and an AC filter (inductor). From there, the current flows through the detector, which can be represented as a kinetic inductance $L_k$ in series with a time-dependent resistance $R_n(t)$. Absorbed photons break the superconducting state, abruptly increasing the resistance, redirecting the current back to the bias tee and across the DC filter, to the signal output. The signal is amplified using both a cryogenic and a room-temperature amplifier. Finally, the signal is measured as a voltage pulse $V(t)$ using an oscilloscope or a counter. Light is guided to the detector via optical fiber centered at \SI{780}{\nano\metre} and \SI{1550}{\nano\metre}. A pulsed laser (\SI{50}{\mega\hertz} repetition rate) is used to characterize the timing jitter of the device, where a correlation measurement is performed between the trigger signal generated by the laser and the detector response.  

Fig. \ref{fig:pulse}b shows a photon detection pulse recorded by our setup as time-dependent voltage response $V(t)$. 
The voltage trace initially follows a conventional exponential decay, however it does not fully relax to \SI{0}{\volt} but remains at a finite voltage due to latching behaviour, whereupon the device reaches a metastable state in which the current-induced Joule self-heating exceeds the electron cooling process. Latching in NbSe$_2$ is especially favoured by the low $R_\square$. Redirection of the current through the parallel resistor speeds up recovery of the superconducting state (see Supplementary Information for details and full-pulse analysis). 


The temporal resolution of SNSPDs is strongly influenced by intrinsic material properties like low kinetic inductance, fast quasi-particle thermalization and crystallinity\cite{korzh_2020}. The pristine quality and the associated properties make NbSe$_2$ a promising candidate for fabricating devices with exceptionally low timing jitter. To provide an initial assessment of the temporal characteristics offered by this novel platform, we measure the timing jitter, or the time distribution of output signals upon photon absorption. Figure \ref{fig:pulse}c shows the histogram of the detection events recorded by the time tagger as function of time. By fitting the data with a gaussian distribution, we report a full-width at half maximum $\sim$45 ps, which sits already within the range of commercially available SNSPD systems, and only an order of magnitude away from the highest temporal resolution ever reported \cite{korzh_2020}, achieved with devices optimized to minimize the propagation delay of the signal into the readout circuit.

\subsection*{Single-photon detection}

\begin{figure*}
\includegraphics[width=1\textwidth]{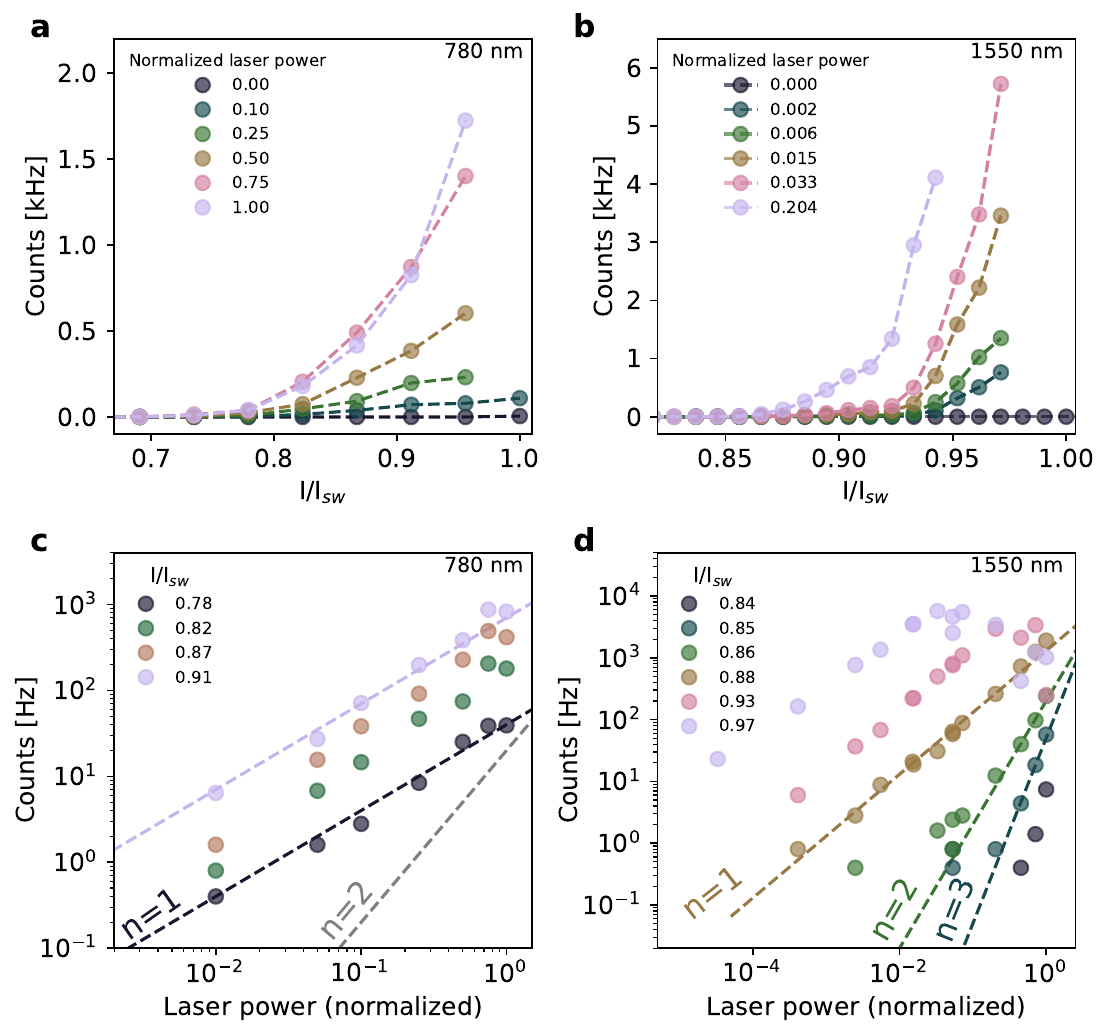}
\caption{\label{fig:spd}$\textbf{Single-photon detection in a superconducting NbSe$_2$ nanowire.}$ Count rates as function of normalized current $I/I_{sw}$ under increasing laser power at $\mathbf{a}$ \SI{780}{\nano\metre} and $\mathbf{b}$ \SI{1550}{\nano\metre}, with counts increasing exponentially until reaching $I_{sw}$, after which the device turns into the resistive state, indicating sub-unity internal detection efficiency. Count rates as function of laser power for several currents at $\mathbf{c}$ \SI{780}{\nano\metre} and $\mathbf{d}$ \SI{1550}{\nano\metre}. Dashed curves indicate power-law behaviour with variable integer exponent $n = 1\text{–}3$. The count rates perfectly follow the dashed curves $n = 1$ over multiple orders of magnitude, indicating single-photon detection.}
\end{figure*}

We continue to focus our attention on the photo-detection behaviour when illuminating the nanowires with CW laser light at \SI{780}{\nano\metre} and \SI{1550}{\nano\metre}. Figures \ref{fig:spd}a and \ref{fig:spd}b show the count rates in the same nanowire device as a function of applied voltage. Below a normalized current threshold $I/I_{sw}\sim$0.75 ($I/I_{sw}\sim$0.85) at \SI{780}{\nano\metre} (\SI{1550}{\nano\metre}) the detector does not record counts. Once we cross the minimum current threshold, the device becomes sensitive showing exponentially increasing counts for increasing current, as expected, up to $>$0.95 $I/I_{sw}$ (0.97 $I/I_{sw}$) at \SI{780}{\nano\metre} (\SI{1550}{\nano\metre}), while maintaining the dark count rate $<$\SI{1}{\hertz}. A detector of unitary internal efficiency would reach and maintain a saturation count rate until reaching the critical voltage turning the NbSe$_2$ from superconducting into resistive. However, our data show a sharp fall from the maximum count rate to $I_{sw}$ without a saturation plateau. The sub-unitary internal detection efficiency can be explained by the interplay between dead time —the bias current recovery time— and cool-down time —the resistive hotspot lifetime. For increasing powers, the cool-down time increases compared to the dead time, making the device more prone to latching. This limits the count rate and decreases the maximal biasing current\cite{kerman_latching}.

To quantify the photon sensitivity, we recall that the probability distribution of detecting $n$ photons from a coherent source when $m$ photons are on average expected is described by the Poisson distribution $\wp=\exp{(-m)}(m^n/n!)$\cite{scully}. If $m$ is sufficiently small\cite{goltsman_2001}, $\lessapprox$0.1, $\wp\simeq(m^n/n!)$, which is a power-law function where $n$ becomes easily identifiable as the slope on the log-log representation of the count rate versus average power. For single-photon detection events, by definition $n$=1, which yields a linear increase of the count rate with increasing photons flux.
Figure \ref{fig:spd}c and Fig. \ref{fig:spd}d show the detector count rates as a function of laser power at \SI{780}{\nano\metre} and \SI{1550}{\nano\metre}, respectively, for several applied biases, compared to curves of discrete power exponent $n$=1,2 and 3. At \SI{780}{\nano\metre}, the count rates exhibits a clear linear increase across two orders of magnitude at all different values of $I/I_{sw}$, confirming single-photon sensitivity. At \SI{1550}{\nano\metre}, starting from the lowest applied biases, initially the count rates increase superlinearly with exponent matching first $n$=3 and then $n$=2. Beyond a current threshold $I/I_{sw}\sim$0.88, the slope reduces to a linear increase over more than three orders of magnitude, unequivocally indicating counting of single-photon events. 

\subsection*{Perspectives for low-energy detection}

\begin{figure*}
\includegraphics[width=1\textwidth]{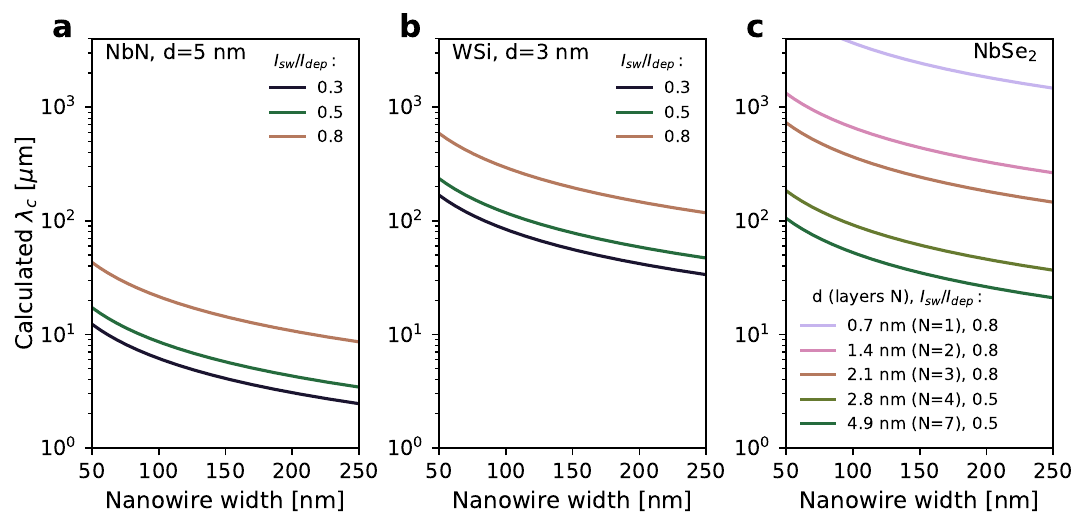}
\caption{\label{fig:cutoff} $\textbf{Predicted cutoff wavelength for representative SNSPD material platforms.}$ $\mathbf{a}$, In first-generation NbN devices, theoretical $\lambda_c$ is limited to few tens of micrometres. $\mathbf{b}$, Second generation amorphous WSi devices extend this value by about an order of magnitude, to beyond \SI{100}{\micro\metre}. $\mathbf{c}$, NbSe$_2$ nanowires of similar geometry are expected to increase $\lambda_c$ by another order of magnitude, breaking the millimetre barrier.}
\end{figure*}

Atomically thin conventional layered superconductors are especially intriguing for detecting low-energy excitations because of single crystal nature and atomically precise thickness. To provide a perspective on this direction, we calculate the cutoff-wavelength $\lambda_c$ of SNSPDs based on their material and device parameters using a simple diffusion-based hot-spot model approximation\cite{Semenov_Engel_Hübers_Il’in_Siegel_2005}, the detection mechanism taking place in the limits of low $D$ and $\tau_{th}$ of NbSe${_2}$ \cite{Vodolazov_2017}:
\begin{equation}
\lambda_c = \frac{hc\zeta}{N_0  \Delta^2 d w \sqrt{\pi D \tau_{th}} } (1-I_{bias}/I_{dep})^{-1}
\end{equation}
with $\zeta$ the conversion efficiency, $\Delta$ the superconducting gap, $N_0$ the density of states at Fermi energy, $D$ the diffusivity, $\tau_{th}$ the thermalization time, $I_{bias}$ and $I_{dep}$ the bias and depairing (the theoretical critical limit) current, $w$ the detector's width, and $d$ the detector's thickness. 
NbSe$_2$ offers especially favourable parameters: low $D$\cite{metuh2025}, extremely fast $\tau_{th}$\cite{wickramaratne_ising_2020}, and sub-nm thickness. 
In Figure \ref{fig:cutoff} we compare $\lambda_c $ for three detectors made of different geometries and materials: a NbN, the workhorse of first-generation SNSPDs; WSi, currently the record-holder for long-wavelength detection\cite{taylor_2023}; and NbSe$_2$ (see Supplementary Information for calculations' details and parameters). Because the current component is strongly affected by the fabrication process, we calculate $\lambda_c$ in a realistic range of $I_{bias}/I_{dep}$.
Figure \ref{fig:cutoff}a shows that NbN is theoretically limited to detecting wavelengths up to a few $\mu$m, as is indeed the case in practice despite $I_{bias}/I_{dep}$ well above 0.5\cite{frasca_determining_2019}. Figure \ref{fig:cutoff}b shows that WSi allows about an order of magnitude improvement to $>$100 $\mu$m, thanks to faster thermalization, lower density of states and lower critical temperature\cite{taylor_2023}. These calculations compare reasonably well with the recent results brushing detection at 30 $\mu$m \cite{taylor_2023} for a device with $I_{bias}/I_{dep}$ $\sim$0.3. Impressively though, Fig. \ref{fig:cutoff}c shows that a four-layer NbSe$_{2}$ matches the $\lambda_c$ of WSi for the same $I_{bias}/I_{dep}$, crosses 1 mm for a two-layer and 4 mm for a single-layer crystal with $I_{bias}/I_{dep}$ $\sim$\SI{0.8}, thus covering the complete teraherz range from 0.03 to 3 mm. Despite the assumptions and approximations of our model, the overall increase in expected $\lambda_c$ provided by NbSe$_2$ is remarkable.

\subsection*{\label{sec:outlook}Conclusion}

Immediate next steps include optimizing detector geometry and architecture to minimize latching and expanding single-photon detection to the MIR ($>$2$ \mu$m).

On the longer run, substrate independence will allow for greater flexibility in device design, enabling seamless incorporation of layered materials' SNSPDs into virtually any dimensionally compatible functional structure. 
Further, their nature and dimensionality will benefit from integration with a broader spectrum of conventional and quantum materials \cite{veyrat_helical_2020, ben_mhenni_breakdown_2025} to enhance device functionality. Collectively, these features form a broad parameter space to explore for the realization of next-generation on-chip integrated quantum photonic circuits as well as detectors for free space optical communication, astronomy, and chemical sensing from the near-IR up to the terahertz range.
Finally, embedding atomically thin superconducting single-excitation detectors into layered materials heterostructures or similar quantum devices may offer a new tool towards local probing of proximal low-energy phases\cite{mhenni_gate-tunable_2024}, adding novel exploratory options for metrology and basic physics' studies.

\section*{Methods}

\subsection*{\label{sec:methods:sample}Sample preparation}
Electrical contacts (Ti/Au $5/20$ nm) were pre-patterned using optical lithography and electron beam evaporation on Si/SiO$_2$ substrates. NbSe$_2$ and hBN flakes (from 2D Semiconductors) were mechanically exfoliated from bulk crystals on PDMS and Si/SiO$_2$ substrates (SiO$_2$ 70 nm thick), respectively. The flakes were selected based on their optical contrast, shape, and cleanliness. hBN crystals are about 20 nm thick, while NbSe$_2$ crystals have a variable thickness from about 10 nm down to 2 layers. The devices were assembled via dry-transfer techniques: hBN was picked up from Si/SiO2 using polycarbonate films\cite{purdie_2018}, followed by NbSe2 picked up from PDMS on the same stamp\cite{castellanos_2014}. Then, the entire heterostructure was released on the pre-patterned contacts.
Heterostructures were then patterned by electron beam lithography, followed by  reactive ion etching with a SF$_6$+Ar gas mixture.

\subsection*{\label{sec:methods:spectroscopy}Photoelectric setup}

We perform cryogenic measurements in a closed-cycle Adiabatic Demagnetization Refrigerator (Kiutra L-type). Photoresponse measurements are performed at \SI{1}{\kelvin}. Bias to the samples is applied with a voltage source $V_b$ (Keithley 2450). Optionally, a resistor in series $R_b$ is applied after the biasing source. The series resistor transforms the voltage bias into a current bias, $I_{bias}$ = $V_b$/$R_b$, stabilizing the current in the circuit. A shunt resistor is applied in parallel, dissipating part of the current when the detector is in the normal conducting state, to facilitate the recovery of superconductivity. The current enters the cryogenic setup and flows through the bias-tee, which consists of a DC filter (capacitor) and an AC filter (inductor). From there, the current gets directed to the detector, acting as an inductance $L_k$ (or kinetic inductance), in series with a time-dependent resistance $R_n(t)$, in an equivalent circuit diagram. Absorbed photons or noise abruptly increase resistance, redirecting the current back to the bias tee and across the AC filter. The signal is amplified using both a cryogenic and a room-temperature amplifier. Finally, the signal is measured as a voltage pulse $V(t)$ using a counter or an oscilloscope.
The sample can be illuminated with two fibre-coupled laser sources, at \SI{780}{\nano\metre} and \SI{1550}{\nano\metre}, whose light is directed into the cryostat to the sample through a fiber attenuator and a beam splitter for the power measurement. One of the two beams is guided to the cryostat, the other to a power meter. The light is collimated to illuminate the sample homogeneously (beam diameter $\sim$2 mm). To characterize the timing jitter of the device, we measure the time correlation between the trigger time of a pulsed laser at \SI{1550}{\nano\metre} (repetition rate \SI{50}{\mega\hertz}) and the SNSPD detection time.

\section*{\label{sec:data_availability}Data availability}
The datasets generated and analyzed during the current study are available from the corresponding authors upon reasonable request.

\section*{\label{sec:acknowledgements}Acknowledgments}
We thank Stefanie Grotowski and Christian Schmid for technical assistance. Work at the Technical Unversity of Munich was supported bythe Germany’s Excellence Strategy (MCQST-EXC-2111, 390814868, e-conversion-EXC-2089), the Deutsche Forschungsgemeinschaft (DFG, German Research Foundation) under projects PQET (INST 95/1654-1) and MQCL (INST 95/1720-1), as well as FI947-8 and SPP 2244 (FI947/7-2). Research groups at the Technical University of Munich also gratefully acknowledge BMFTR for financial support via SPINNING (13N16214), QR-N (16KIS2197) and PhotonQ (13N15760), the Munich Quantum Valley supported by the State of Bavaria, and the German Federal Ministry of Education and Research via the funding program quantum technologies - from basic research to market (contract numbers QPIS: 16K1SQ033  and QPIC-1: 13N15855). B.M. acknowledges support from the European Research Council (ERC-StG ``TuneTMD", grant no. 101076437) and Villum Fonden (project no. VIL53033).


\section*{\label{sec:contributions}Author contributions}
L.Z., K.M., J.J.F., and M.B. conceived and managed the research. L.Z., Al.P., B.S., An.P., and F.W. fabricated the devices. L.Z., Al.P., and B.S. performed the optical measurements. L.Z., Al.P., and M.B. analyzed the results. J.J.F. and K.M. obtained third-party funding and provided experimental and nanofabrication infrastructure. 
All authors discussed the results. L.Z. and M.B. wrote the manuscript with input from all authors.\newline

\section*{\label{sec:interests}Competing interests}
The authors declare no competing interests.


\newpage

\beginsupplement
\maybeonecolumn

\section*{\label{sec:si}Supplementary information}
Superconducting transitions of patterned devices; Details on detection pulse and latching behaviour; Effect of laser illumination on the I-V curves and switching current; Oscilloscope time trace of detection events; Details and parameters on cutoff-wavelength calculation. 

\subsection{\label{transitions}Superconducting transitions of patterned devices}

\begin{figure*}
\includegraphics[width=0.8\textwidth]{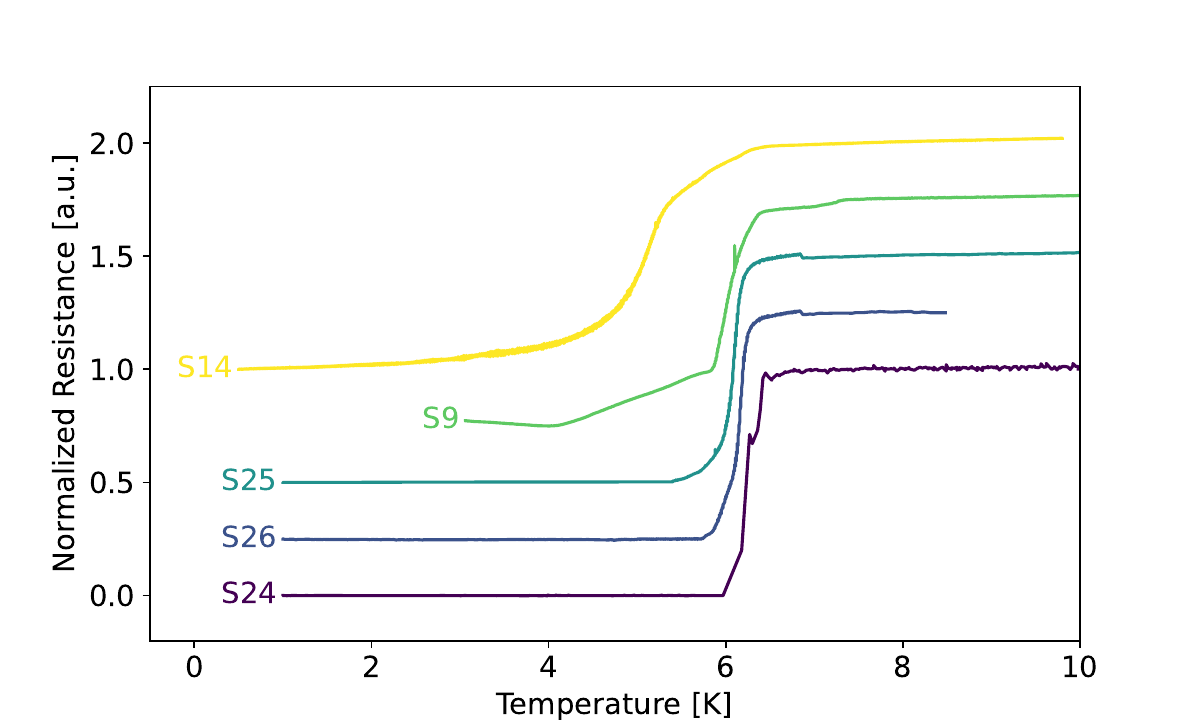}
\caption{\label{fig:transitions} $\textbf{Superconducting transitions of patterned devices.}$}
\end{figure*}

In Figure S\ref{fig:transitions}, the resistive transition for several patterned devices is shown. The values of resistance are normalized, removing the contact resistance, and are shifted for clarity. The transition temperatures varies based on the number of NbSe$_2$ layers in the device, ranging between 5K and 6.5K. The samples presented have the following number of layers: S14 - 2 layers; S9 - 3 layers; S25 - 7 layers; S26 - 5 layers; S24 - bulk.\\

\subsection{\label{extended_pulse}Extended pulse}

\begin{figure*}
\includegraphics[width=0.8\textwidth]{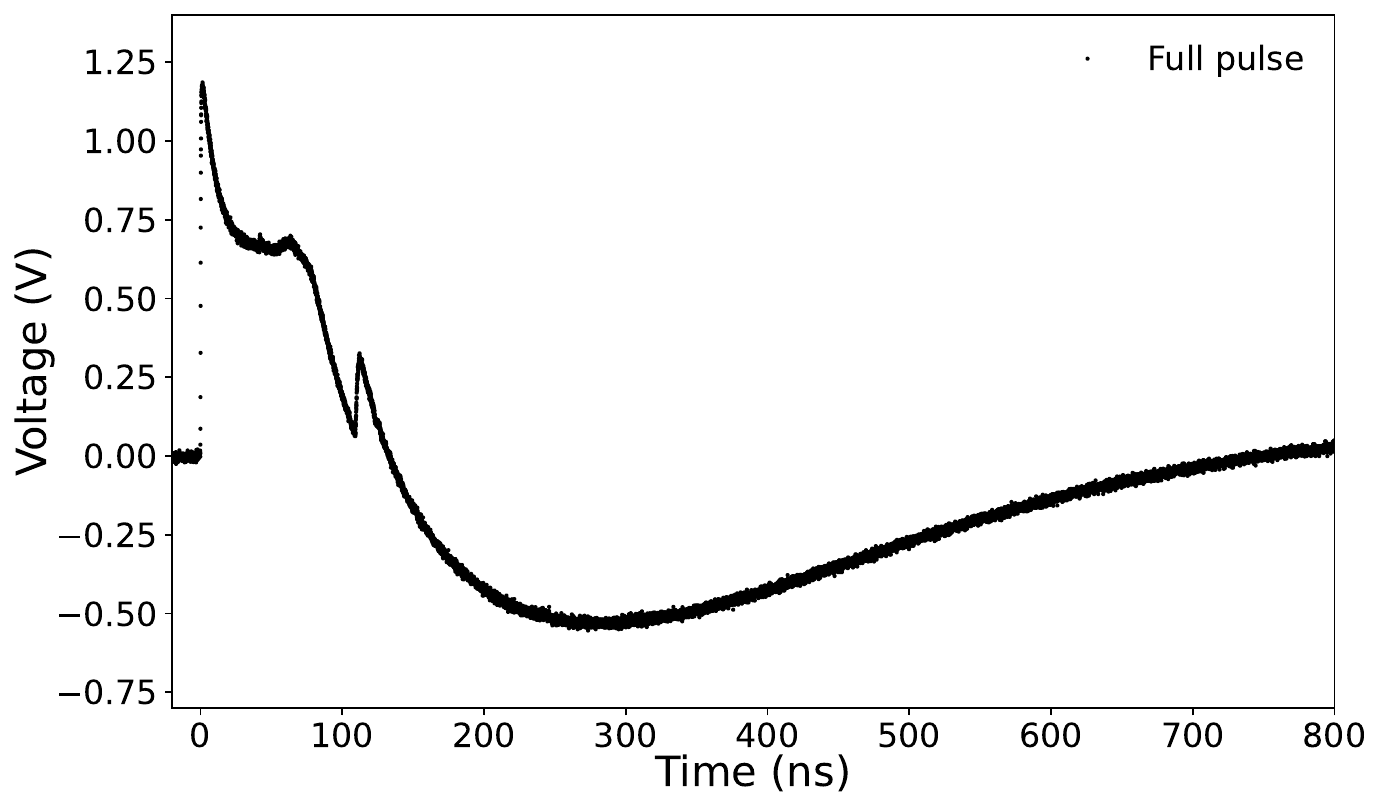}
\caption{\label{fig:extended_pulse} $\textbf{Extended detection pulse.}$}
\end{figure*}

The pulse initially decays exponentially before reaching the latching plateau from $\sim$20 ns to $\sim$100 ns. Around 110 ns a reflection from the wiring causes a spike. After that, the pulse undershoots below 0 V due to the instrumental effect of the cryogenic amplifier, until it fully recovers $\sim$800 ns.\\
The latching behaviour is mainly governed by the kinetic inductance of the device, which can be estimated by $L_k = \frac{\hbar R}{\pi \Delta} = \frac{\hbar R}{1.764 \pi k_B T_c}$ \cite{tinkham2004introduction}, where $R$ is the resistance of the device above the superconducting transition, and $T_c$ is the critical temperature. The device measured here has $L_k \approx 319$ pH. Given the low value of $L_k$, current is restored in the device before complete thermalization, thus before restoring the superconducting state, and therefore it results in a finite resistance. The superconducting state is then recovered (transition from non-zero voltage to 0) due to the implementation of the parallel resistor at room temperature. The fact that the current in the device is not immediately redirected to the parallel resistor comes from the circuit configuration implemented, given that the AC filter is in between the device and the shunt. 

\subsection{\label{current_vs_illumination}Effect of light illumination on switching current}
\begin{figure*}[h]
\includegraphics[width=0.6\textwidth]{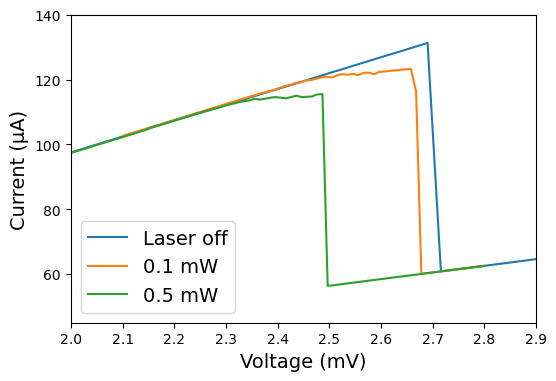}
\caption{\label{fig:curr_v_ill} $\textbf{I-V curves under light illumination.}$}
\end{figure*}

The I-V curves for the device are measured under illumination with different 1550 nm laser powers. We notice a change in the switching current depending on the applied laser power. Increasing the laser power, we register a decrease of I$_{sw}$ - the sudden drop in current in the  I-V sweep - as expected due to the higher photon flux.\\
During illumination, the I-V curve do not present a sharp transition to the normal conducting state. This can be explained by the latching state of the device, as explained in Section EXTENDED PULSE.

\subsection{\label{pulse_train}Pulse train}

\begin{figure*}[h]
\includegraphics[width=0.8\textwidth]{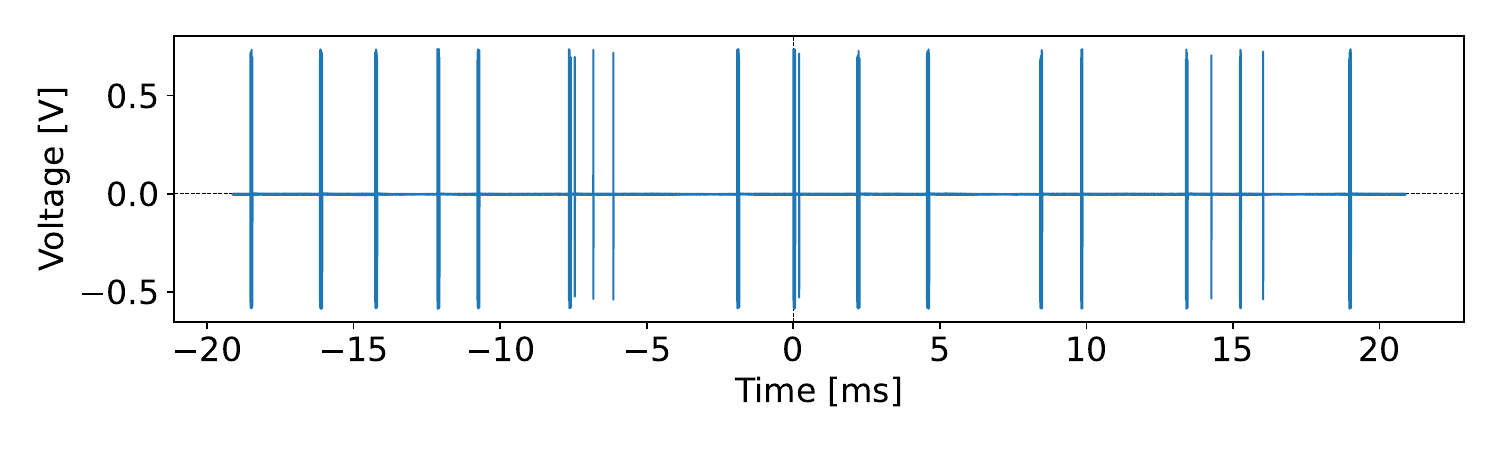}
\caption{\label{fig:pulse_train} $\textbf{Detection pulses under illumination.}$}
\end{figure*}

The time trace recorded with an oscilloscope is shown in Figure S\ref{fig:pulse_train}. The time window has a length of 50 ms, where several pulses, as the one presented above are recorded.

\subsection{\label{sec:methods:cutoff_calculations} Cutoff wavelength efficiency calculations}

\begin{table*}[ht]
\centering
\caption{Material parameters used in the diffusion-based hot-spot model.}
\resizebox{\textwidth}{!}{%
\begin{tabular}{|l|c|c|c|}
\hline
\textbf{Parameter} & \textbf{NbSe\textsubscript{2}} & \textbf{WSi} & \textbf{NbN} \\
\hline
$N_0$ [$10^{47}$ m$^{-3}$J$^{-1}$] & 1.3\cite{wickramaratne_ising_2020} & 1.08\cite{taylor_2023} & 1.6\cite{Semenov_Engel_Hübers_Il’in_Siegel_2005} \\
$\Delta = k_B T_c$ [meV] & 0.43 (T$_c$ = 5 K) & 0.112 (T$_c$ = 1.3 K) & 0.43 (T$_c$ = 5 K) \\
$D$ [m$^2$s$^{-1}$] & $1\times10^{-5}$ \cite{metuh2025} & $5.3\times10^{-5}$ \cite{taylor_2023} & $3.5\times10^{-5}$ \cite{Semenov_Engel_Hübers_Il’in_Siegel_2005} \\
$\tau_{th}$ [ps] & 1\cite{anikin_ultrafast_2020} & 66\cite{zhang_superconducting_2018, sidorova_nonbolometric_2018} & 14\cite{Semenov_Engel_Hübers_Il’in_Siegel_2005, sidorova_nonbolometric_2018} \\
$d$ [nm] & 0.7-bulk & 3\cite{taylor_2023} & 5\cite{Semenov_Engel_Hübers_Il’in_Siegel_2005} 
\\
\hline
\end{tabular}
}
\label{tab:material_params}
\end{table*}

\begin{table*}[ht]
\centering
\caption{Thickness dependent critical temperature for NbSe$_2$.}
\resizebox{\textwidth}{!}{%
\begin{tabular}{|l|c|c|c|}
\hline
\textbf{Layers} & \textbf{Cao \textit{et al.} \cite{cao_2015}} & \textbf{Xi \textit{et al.} \cite{xi_ising_2016}} & \textbf{This work} \\
\hline
1 [K] & 1.95 & 3.1 & - \\
2 [K] & 4.5 & 5.4 & 5.1 \\
3 [K] & 5.5 & 5.9 & 5.6 \\
4 [K] & - & 6.2 & - \\

7 [K] & 6.5 & 6.9 & 6.0 \\
\hline
\end{tabular}
}
\label{tab:critical_temp}
\end{table*}

To calculate the cutoff wavelength according to the diffusion-based hot-spot model described in the main text, we employ materials' parameters provided in Table S\ref{tab:material_params}. 
We adopt a $\zeta$=1 for all materials to adopt a prudent approach, although due to lower carrier density, NbSe$_2$ is expected to have a higher $\zeta$ than conventional superconductors \cite{di_battista_2024}.
We calculate $N_0^{NbSe_2}$=1.3$\times10^{47} m^{-3}J{-1}$ from 2.7 states per unit cell\cite{wickramaratne_ising_2020} with cell volume V$_{cell}$=$(\sqrt{3}/2)a^2 c$, a=3.45 $\mathring{A}$, c=12.550 $\mathring{A}$.
We calculate $\Delta$=$k_BT_c$, with $k_B$ the Boltzmann constant, from the superconducting transition temperatures  $T_c^{WSi}$=1.3 K\cite{taylor_2023}, $T_c^{NbN}$=5 K \cite{Semenov_Engel_Hübers_Il’in_Siegel_2005}, and $T_c^{NbSe_2}$ varies based on the number of layers in the device, following values from literature and from this work presented in Table S\ref{tab:critical_temp}\cite{cao_2015, xi_ising_2016}.
We calculate $D^{NbSe_2}=$ $1\times10^{-5}$ $m^2s^{-1}$ from magnetic field dependence data of the superconducting transition in a similar device \cite{metuh2025}. 

Finally, $\tau_{th}$ is mainly influenced by the electron-phonon scattering time, which is the value we use. In the case of NbN the value is 14 ps\cite{Semenov_Engel_Hübers_Il’in_Siegel_2005, sidorova_nonbolometric_2018}. For few-nm-thick amorphous WSi films, the amorphous nature of the material increases the electron-phonon scattering time to$\sim$66 ps \cite{zhang_superconducting_2018, sidorova_nonbolometric_2018}. In NbSe$_2$, this value falls to a much shorter at $\sim$1 ps \cite{anikin_ultrafast_2020}.

\maybetwocolumn

\section{Bibliography}

\input{output.bbl}


\end{document}

%% file: output.bbl
%